\documentclass{pasj01}
\Received{$\langle$reception date$\rangle$}
\Accepted{$\langle$acception date$\rangle$}
\Published{$\langle$publication date$\rangle$}

\usepackage{color}
\usepackage{bm}
\usepackage[switch,mathlines]{lineno}

\begin{document}

\title{Formation of massive star clusters with and without iron abundance spreads in a dwarf galaxy merger}

\author{Hidenori Matsui${}^{1,2}$}
\author{Kenji Bekki${}^{2}$}
\author{Madeleine McKenzie${}^{3,4}$}
\author{Takayuki R. Saitoh${}^{5,6}$}
\altaffiltext{1}{National Institute of Technology, Asahikawa College, Shunkodai 2-2-1-6, Asahikawa, Hokkaido, 071-8142, Japan}
\altaffiltext{2}{International Centre for Radio Astronomy Research (ICRAR), M468, The University of Western Australia, 35 Stirling Hwy, Crawley, Western Australia, 6009, Australia}
\altaffiltext{3}{Research School of Astronomy \& Astrophysics, Australian National University, Canberra, ACT 2611, Australia}
\altaffiltext{4}{ARC Centre of Excellence for Astrophysics in Three Dimensions (ASTRO-3D), Canberra 2611, Australia}
\altaffiltext{5}{Department of Planetology, Graduate School of Science, Kobe University, 1-1 Rokkodai-cho, Nada-ku, Kobe, Hyogo 657-8501, Japan}
\altaffiltext{6}{Center for Planetary Science (CPS), Graduate School of Science, Kobe University, 1-1 Rokkodai-cho, Nada-ku, Kobe, Hyogo 657-8501, Japan}
\email{matsui@asahikawa-nct.ac.jp}


\maketitle

\begin{abstract}
To study the formation of star clusters and their properties in a dwarf-dwarf merging galaxy, we have performed a numerical simulation of a dwarf-dwarf galaxy merger by using the Tree+GRAPE $N$-body/SPH code ASURA.
In our simulation, 13 young star clusters are formed during the merger process.
We show that our simulated star clusters can be divided into two types: with and without [Fe/H] abundance variations.
The former is created by a seed star cluster (the first-generation stars) formed in compressed gas. These stars contaminate the surrounding gas by Type II supernovae (SNe).
At that time, the energy injection is insufficient to induce an outflow of the surrounding gas.
After that, the contaminated gas falls into the seed, thereby forming a new generation of stars from the contaminated gas.
We also show that most star clusters are formed in the galactic central region after the second encounter and fall into the galactic center due to dynamical friction within several hundred Myr.
As a result, close encounters and mergers between the clusters take place.
Although the clusters with shallower gravitational potential are tidally disrupted by these close encounters, others survive and finally merge at the center of the merged dwarf galaxies to create a nuclear star cluster. Therefore, the nuclear star cluster is comprised of various stellar components in [Fe/H] abundance and age.
We discuss our work in the context of observations and demonstrate the diagnostic power of high-resolution simulations in the context of star cluster formation.
\end{abstract}


\section{Introduction}\label{sec:intro}

Observations have revealed the peculiar properties of globular clusters (GCs; recent reviews include \citet{Gratton+2012,2018ARA&A..56...83B,Gratton+2019,2022Univ....8..359M}).
It is well understood that Galactic and extragalactic GCs are generally composed of multiple stellar populations (e.g., \citet{Norris_Freeman1979,2014ApJ...797...15L,2019MNRAS.487.3815M}).
Additionally, the anomalous chemical composition of light element abundances, such as a correlation in Na-N \citep{Nataf+2019} and an anti-correlation in Na-O \citep{2008ApJ...672L..29Y,Carretta+2009a,Carretta+2009b}, are a fundamental characteristic of GCs.

Based on photometric observations, GCs are typically divided into two types; Type I and II \citep{Milone+2017}.
Type I clusters are “classical” GCs which exhibit a bimodal distribution in light element abundances and are homogeneous in iron at the limit of typical spectroscopic uncertainties (${\rm [Fe/H]} \sim 0.05$; \citet{Carretta+2009c}).
These clusters are composed of a population of stars chemically similar to Milky Way halo stars (the first population (1P)), and a population enhanced in elements such as N, Na and Al (the second population (2P)).
In contrast, Type II clusters exhibit two or more populations differing in their light element abundances, and are accompanied by iron peak and neutron capture element abundance variations.
This assumes some contribution from supernovae in order to produce iron peak element variations, however, the source of neutron capture variations remains uncertain.
Prominent examples of Type II clusters include but are not limited to $\omega$ Centauri ($\omega$ Cen) \citep{Nitschai+2023}, M22 \citep{McKenzie+2022} and NGC6934 \citep{2018ApJ...859...81M}.
The differences in origins between Type I and II clusters are still a matter of debate.

Galaxy mergers are an interesting test case for studying the formation, evolution, and properties of star clusters.
Previous observations indicate that merging galaxies produce star clusters effectively like in nearby galaxies such as the large and small magellanic clouds \citep{Williams+2022}, NGC4038/4039 \citep{1995AJ....109..960W,2008A&A...489.1091M}, NGC4676 \citep{2007ApJ...660L.105C}, and NGC3256 \citep{2007ApJ...658..993T}, young massive star clusters with $10^{5-6}~{\rm M}_{\odot}$ are formed during the merging process.
In merger remnants, star clusters expected to be formed during the interaction have also been observed \citep{2022MNRAS.511..393E}. 

Reproducing the formations of star clusters through a galaxy merger using numerical simulations is difficult due to the limited mass and spatial resolutions.
However recent high resolution numerical simulations of gas rich merging galaxies, which realize the multi-phase nature of interstellar medium, take important steps towards overcoming these difficulties and reproduce the formation of star clusters during galaxy-galaxy merger \citep{2009PASJ...61..481S,2012ApJ...746...26M,2013MNRAS.430.1901H,2015MNRAS.446.2038R}.
\citet{2019PASJ...71...19M} analyzed the properties of star clusters formed via a merger process, revealing that star clusters composed of 1P stars captured gas when they passed through a dense gas region.
In the cluster, 2P stars form from the captured gas and as a result, star clusters with multiple stellar populations are created.
The captured gas is expected to be contaminated by Type II supernovae (SNe) as a result of a starburst, as such, it is likely that the resulting clusters would be classified as Type II GCs due to the presence of iron abundance variations.

\citet{2019PASJ...71...19M} focused on a merger between massive disk galaxies.
However, observations have shown evidence of a merger of not only massive galaxies but also dwarf galaxies \citep{2018ApJS..237...36P}.
The blue compact dwarf galaxy VCC 848 is one such dwarf-dwarf merging galaxy.
In VCC 848, abundant gas and young star clusters have been observed \citep{2020ApJ...900..152Z}, which suggests that dwarf-dwarf merging galaxies are an ideal test bed for investigating the formation of star clusters.
Previous simulations have revealed that massive star clusters can be formed in the dwarf-dwarf galaxy merger leading to the blue compact galaxy \citep{2008MNRAS.388L..10B}.
However, the chemical abundance patterns of the massive star clusters have not been investigated by these previous simulations in detail.
It is thus worthwhile to investigate whether Type II GCs with significant ${\rm [Fe/H]}$ spreads can be formed in dwarf-dwarf merging process.
Recently, \citep{2024arXiv240209518L} have performed simulations of a dwarf galaxy merger.
They have revealed that stellar winds from massive stars trigger self metal-enrichment of star clusters.

In this paper, we performed a high resolution simulation of a dwarf-dwarf merging galaxy to study evolution and properties of the resulting star clusters.
Here, our numerical simulation realizes the multi-phase nature of interstellar medium (ISM), which allows us to understand the formation and evolution of star clusters.
Our numerical method, results, discussions, and conclusion are described in \S 2, \S 3, \S 4, and \S 5, respectively.

\section{Methods} \label{sec:method}

\subsection{Simulation models} \label{subsec:model}

We perform a numerical simulation of a gas rich merger of two dwarf galaxies.
The two galaxies are same models; each initially consisting of a gas disk, a stellar disk, and a halo.
We assume the gas and the stellar disks to be an exponential disk.
The masses of the gas and the stellar disks are $2.5\times 10^8~{\rm M}_{\odot}$.
The scale radii of the gas and the stellar disks are $1.5~{\rm kpc}$ and $0.7~{\rm kpc}$, respectively.
The density profile of the halo is assumed to be the NFW profile \citep{1996ApJ...462..563N}.
The mass and scale radius are set to be $5.0\times 10^{10}~{\rm M}_{\odot}$ and $8.0~{\rm kpc}$, respectively.
The halo contains halo gas with $1.0\times 10^8~{\rm M}_{\odot}$ which corresponds to $0.2~\%$ of the total halo mass budget.
In our dwarf galaxy model, the baryon fraction to the total halo mass is comparable with observational predictions \citep{2019A&A...626A..56P}.
On the basis of the above models, initial distributions of smoothed particle hydrodynamics (SPH) and $N$-body particles are generated by the software MAGI \citep{2018MNRAS.475.2269M}.
The initial temperatures of the gas disk and the halo gas are $10^4~{\rm K}$ and $10^5~{\rm K}$, respectively.
The initial gas has metallicity abundance with ${\rm [Fe/H]} = -1.6$ which is consistent with observations of local dwarf galaxies \citep{2013ApJ...779..102K}.

Initial positions of the galactic centers of two galaxies are set to be $(x~[{\rm kpc}],y~[{\rm kpc}],z~[{\rm kpc}]) = (-25,0,0)$ and $(25,0,0)$, respectively.
Their initial velocities are $(v_x~[{\rm km/s}],v_y~[{\rm km/s}],v_z~[{\rm km/s}]) = (0,-9.2,0)$ and $(0,9.2,0)$, respectively.
This setups results in elliptical orbits for dwarf galaxy pair with pericenter distance of $2.5~{\rm kpc}$.
The initial normalized spin axes of two disks are set to be $(x,y,z)=(0.35,0.35,0.87)$ and $(-0.35,-0.35,0.87)$, respectively.

\subsection{Simulation methods} \label{subsec:method}

We perform a numerical simulation of a merger of two dwarf galaxies using Tree+GRAPE $N$-body/SPH code ``ASURA'' \citep{2008PASJ...60..667S}.
ASURA takes into account a wide temperature range of radiative cooling ($10~{\rm K}<T<10^8~{\rm K}$), star formations from cold and dense gas, and gas heating by SN feedback.
The cooling function depends on the metallicity.
Star formation takes place when an SPH particle satisfies the following conditions: (1) $T<100~{\rm K}$, (2) $n_{\rm H} >100~{\rm cm^{-3}}$, and (3) $\nabla \cdot \bm{v} < 0$,
where $T$, $n_{\rm H}$, and $\bm{v}$ are the temperature, number density, and gas velocity, respectively.
An SPH particle fulfilling these criteria spawns a star particle with a mass equal to one-third of an SPH particle (i.e., $2\times 10^3 \times 1/3 = 6.67\times 10^2~{\rm M}_{\odot}$, see below).
A newly formed star particle is treated as a single stellar population (SSP) which we assume follows a Kroupa initial mass function \citep{2001MNRAS.322..231K}.
Due to Type II SNe, the SSP particles distribute thermal energy to the surrounding SPH particles.
The energy per SN is $10^{51}~{\rm erg}$.
This treatment of the interstellar medium (ISM) emulates star cluster formations naturally without any prescriptions \citep{2009PASJ...61..481S,2012ApJ...746...26M}.

Metal contamination of the gas is taken into account using the Chemical Evolution Library (CELib) \citep{2017AJ....153...85S}.
In the timescale of our simulation, gas is contaminated mainly by Type II SNe.
We adopt the yield tables of \citet{2013ARA&A..51..457N} for Type II SNe.
Type Ia SNe, asymptotic giant branch (AGB) stars, and neutron star mergers (NSMs) are also taken into account.
We adopt the yield tables from \citet{2013MNRAS.429.1156S} for Type Ia, a combination of \citet{2008A&A...490..769C}, \citet{2010MNRAS.403.1413K}, \citet{2013A&A...557A.106G}, and \citet{2014MNRAS.437..195D} for AGB stars, and \citet{2014ApJ...789L..39W} for NSMs.
Metal diffusion of $0.01$ is also taken into account \citep{2017ApJ...838L..23H}.

In our simulation, the numbers of SPH, old star, and dark matter particles are $3.5\times 10^5$, $2.5\times 10^5$, and $4.99\times 10^7$, respectively.
Here, the old star particles constitute stellar disks, and we do not take into account stellar evolution of these particles. 
The mass of all type particles except for newly formed star particles is $2\times 10^3~{\rm M}_{\odot}$.
The gravitational softening length is $1~{\rm pc}$ for all particles.
Since these mass and spatial resolutions are much higher than our previous studies \citep{2012ApJ...746...26M,2019PASJ...71...19M}, this simulation has a sufficient resolution to follow the formation and evolution of star clusters in a galaxy-galaxy merging process.
In the context of the wider literature of theoretical studies of dwarf galaxy simulations, we simulate two, much bigger dwarf galaxies than in \citet{2022MNRAS.511.5672R} at a comparable spatial resolution, however, we do not run our simulations for cosmological time as they have.
As another recent example, \citet{2024MNRAS.529.4104V} also investigated the formation of nuclear star clusters within a cosmological simulation which stochastically creates star particles with an initial mass of $1.026\times 10^3~{\rm M}_{\odot}$, again comparable to our new star mass resolution of $6.67\times 10^2~{\rm M}_{\odot}$.

\subsection{Identification of star clusters} \label{subsec:sc}

To identify star clusters with our simulation data, we first calculate the gravitational potential.
Secondly, we pick up newly formed star particles within $20~{\rm pc}$ of the star particle with a local minimum potential.
When the total mass around this local potential exceeds $10^5~{\rm M}_{\odot}$, we regard the stellar group as a star cluster.
Since the mass of young star particles is $\sim 7\times 10^2~{\rm M}_{\odot}$, the star clusters are comprised of at least $\sim 100$ star particles.

\section{Results} \label{sec:results}

\subsection{Formations of star clusters in a merging galaxy} \label{sec:merger}

\begin{figure}
 \begin{center}
  \includegraphics[width=80mm]{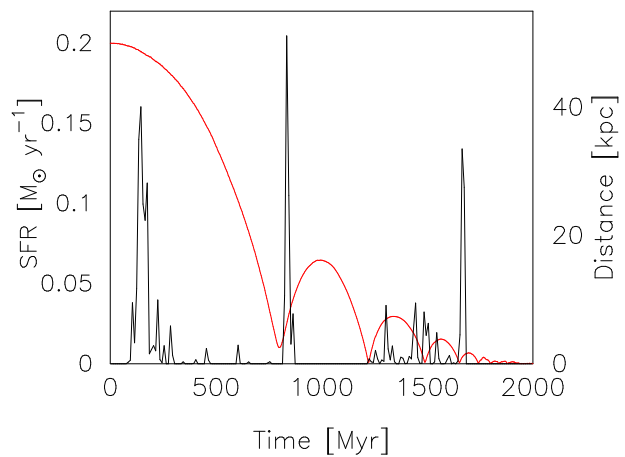}
 \end{center}
 \caption{
  Time evolution of the SFR and distance of two galaxies.
  The black and red lines show SFR and distance of two galactic centers, respectively.
  The centers of galaxies are decided by the potential minimums in the dark halos.
 }
 \label{sfr}
\end{figure}

Figure~\ref{sfr} shows time evolution of the distance between two galaxies and star formation rate (SFR).
Before the first encounter of the two dwarf galaxies, active star formations take place in each disk after around $150~{\rm Myr}$ due to self-gravitational instability.
The SFR of each galaxy's disk peaks before the first encounter at $\sim 0.15~{\rm M}_{\odot}~{\rm yr}^{-1}$.
Following this, star formation is reduced: the average SFR of the two galaxies after this peak and before the first encounter is $0.01~{\rm M}_{\odot}~{\rm yr}^{-1}$.

The first encounter where the two gas disks collide takes place at $\sim 800~{\rm Myr}$, and its pericenter distance is $2.5~{\rm kpc}$.
The collision compresses gas and produces a large gas filament structure at the collision interface as observed in interacting galaxies \citep{2018ApJ...860L..14K}.
In the filament, shock-induced star formation takes place similarly to \citet{2009PASJ...61..481S}, thereby increasing the SFR to $0.2~{\rm M}_{\odot}~{\rm yr}^{-1}$ which is comparable to the SFR of local starburst dwarf galaxies \citep{2016ApJ...817...20M}.
Active star formation is accompanied by the formation of three star clusters.
The second, third, and fourth encounters occur at $\sim 1.2~{\rm Gyr}$, $\sim 1.5~{\rm Gyr}$, and $\sim 1.65~{\rm Gyr}$, respectively.
The gas compression resulting from these encounters also triggers active star and star cluster formation, thereby increasing the SFR from $0.05$ to $0.2~{\rm M}_{\odot}~{\rm yr}^{-1}$.
The second encounter creates two additional star clusters, the third creates three, and the fourth encounter produces four more clusters.

The fourth encounter induces strong star formation in a compact region of the galaxies rather than in a widespread region like in previous encounters.
As a result, strong energy injection from Type II SNe results in a radial outflow of a large amount of gas since the gravitational potential is not deep enough to bind such gas to the merging dwarfs.
Due to the outflow, little gas remains in the merger remnant.
Consequently, star and star cluster formation is quenched due to gas depletion.
The galactic cores finally merge after $\sim 1.8~{\rm Gyr}$.

\begin{figure}
 \begin{center}
  \includegraphics[width=80mm]{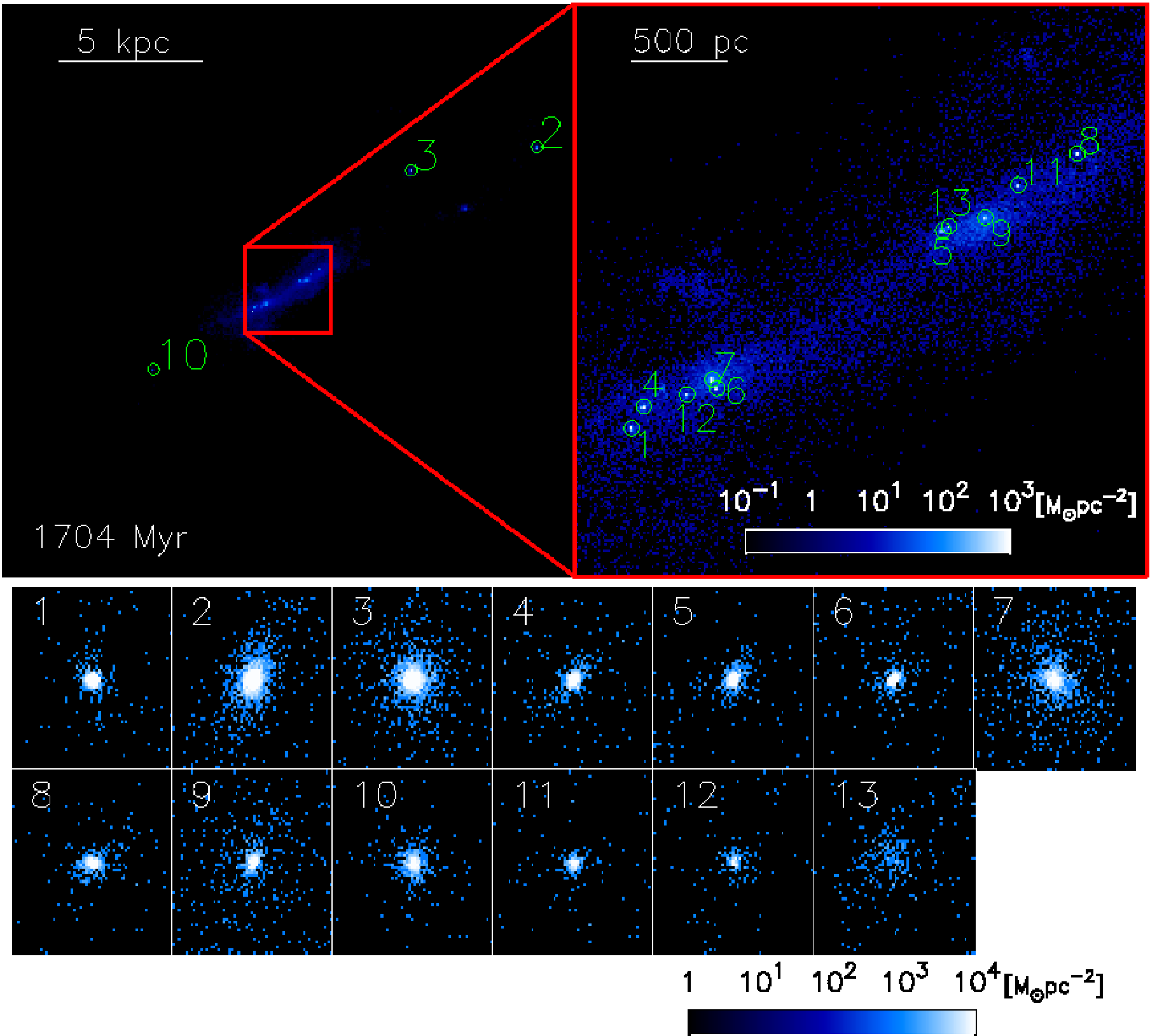}
 \end{center}
\caption{
 Stellar distribution map (upper panels) and star clusters (bottom panels).
 The upper right panel zooms in on the square region in the upper left panel.
 In these panels, green circles show positions of star clusters.
 Cluster IDs are attached to the circles.
 The sizes of the upper left and right panels are $20~{\rm kpc}\times 20~{\rm kpc}$ and $300~{\rm pc}\times 300~{\rm pc}$, respectively.
 The 13 lower panels show individual star clusters.
 The size of each lower panel is $50~{\rm pc}\times 50~{\rm pc}$.
}
\label{stellar_map}
\end{figure}

Figure~\ref{stellar_map} shows the distribution of newly formed stars after $1.704~{\rm Gyr}$.
In this figure, identified star clusters are circled in green and are assigned IDs in order of the depth of their gravitational potentials.
Their mass, formation time, half mass radius, minimum potential, the standard deviation in [Fe/H], and the type of a star cluster are summarized in Table~\ref{tab:clusters}.
Star clusters with IDs 2, 3, and 10 form during the first encounter.
Star clusters with IDs 5 and 13, IDs 8, 9, and 11, and IDs 1, 4, 6, and 12 form at the second, the third, and the fourth encounters, respectively.
Only the star cluster with ID 7 forms in the disk before the merger.

\begin{table*}
  \tbl{Properties of star clusters.}{%
  \begin{tabular}{cccccccc}
      \hline
	ID	&	Mass [${\rm M}_{\odot}$]\footnotemark[a]	&	Formation time [${\rm Myr}$]\footnotemark[b]	&	$r_h$ [${\rm pc}$]\footnotemark[c]	&	Minimum potential [${\rm erg/g}$]\footnotemark[d]	&	Median ${\rm [Fe/H]}$\footnotemark[e]		&	$\sigma$ in [Fe/H]\footnotemark[f]	& Type\footnotemark[g]	\\
      \hline
	1	&	$1.75\times 10^6$	&	$1669$	&	$0.8$	&	$-5.6\times 10^{13}$	&	$-0.95$	&	$0.03$	&	Type I	\\
	2	&	$1.28\times 10^6$	&	$835$	&	$2.5$	&	$-2.2\times 10^{13}$	&	$-1.30$	&	$0.14$	&	Type II	\\
	3	&	$1.13\times 10^6$	&	$824$	&	$2.6$	&	$-1.9\times 10^{13}$	&	$-1.16$	&	$0.21$	&	Type II	\\
	4	&	$4.75\times 10^5$	&	$1660$	&	$1.6$	&	$-1.1\times 10^{13}$	&	$-0.98$	&	$0.01$	&	Type I	\\
	5	&	$3.80\times 10^5$	&	$1300$	&	$1.6$	&	$-8.6\times 10^{12}$	&	$-1.16$	&	$0.04$	&	Type I	\\
	6	&	$3.01\times 10^5$	&	$1666$	&	$1.4$	&	$-7.1\times 10^{12}$	&	$-0.95$	&	$0.09$	&	Type I	\\
	7	&	$5.06\times 10^5$	&	$144$	&	$3.9$	&	$-6.6\times 10^{12}$	&	$-1.03$	&	$0.44$	&	Type II	\\
	8	&	$3.05\times 10^5$	&	$1499$	&	$1.9$	&	$-6.3\times 10^{12}$	&	$-1.04$	&	$0.01$	&	Type I	\\
	9	&	$2.97\times 10^5$	&	$1483$	&	$2.4$	&	$-5.6\times 10^{12}$	&	$-1.05$	&	$0.07$	&	Type I	\\
	10	&	$2.84\times 10^5$	&	$862$	&	$2.7$	&	$-4.5\times 10^{12}$	&	$-1.35$	&	$0.06$	&	Type I	\\
	11	&	$1.40\times 10^5$	&	$1543$	&	$1.7$	&	$-3.1\times 10^{12}$ &	$-0.99$	&	$0.03$	&	Type I	\\
	12	&	$1.03\times 10^5$	&	$1670$	&	$2.4$	&	$-1.9\times 10^{12}$	&	$-0.95$	&	$0.32$	&	Type I	\\
	13	&	$1.08\times 10^5$	&	$1432$	&	$7.9$	&	$-7.0\times 10^{11}$	&	$-1.11$	&	$0.05$	&	Type I	\\
    \hline
    \end{tabular}}\label{tab:clusters}
\begin{tabnote}
\footnotemark[a] Mass of a star cluster.  \\ 
\footnotemark[b] Formation time defined by the time that total stellar mass in the cluster reaches $10~\%$ of the cluster mass.\\
\footnotemark[c]  Half-mass radius of a star cluster. \\ 
\footnotemark[d]  Minimum gravitational potential of the star particle in a star cluster. \\
\footnotemark[e] The median of ${\rm [Fe/H]}$ distribution in a star cluster. \\
\footnotemark[f] The standard deviation of [Fe/H] in a star cluster. \\
\footnotemark[g] Type of a star cluster. When the $\sigma$ is higher than $0.1$, type of star clusters is classified as Type II. Although the star cluster with ID 12 has $\sigma =0.32$, we regard it as Type I.
This is because such high $\sigma$ in this cluster is attributed to a minor fraction of star particles with high [Fe/H]. \\
\end{tabnote}
\end{table*}

\subsection{${\rm [Fe/H]}$ distributions in each star cluster} \label{sec:first_cluster}

\begin{figure}
 \begin{center}
  \includegraphics[width=80mm]{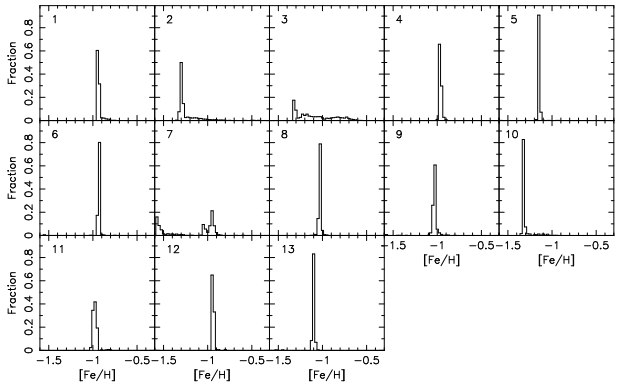}
 \end{center}
\caption{
 Distribution of stellar populations of each star cluster in ${\rm [Fe/H]}$.
 The upper left number in each panel corresponds to star cluster ID in Figure~\ref{stellar_map}.
 The bin width is $0.25~{\rm dex}$.
}
\label{feh_distribution}
\end{figure}

Figure~\ref{feh_distribution} shows ${\rm [Fe/H]}$ distributions of stellar populations in each star cluster.
Most star clusters have a single peak in their [Fe/H] distributions, of which the width is less than $0.05~{\rm dex}$.
The peak value of [Fe/H] strongly depends on the chemical evolution of the host galaxy and the formation time of the stars since the metallicity of the ISM increases with time.
The clusters formed at the first, the second, the third, and the forth encounters roughly have the peak with ${\rm [Fe/H]}\sim -1.3$, $\sim -1.1$, $\sim -1.0$, and $-0.9$, respectively.

\begin{figure}
 \begin{center}
  \includegraphics[width=80mm]{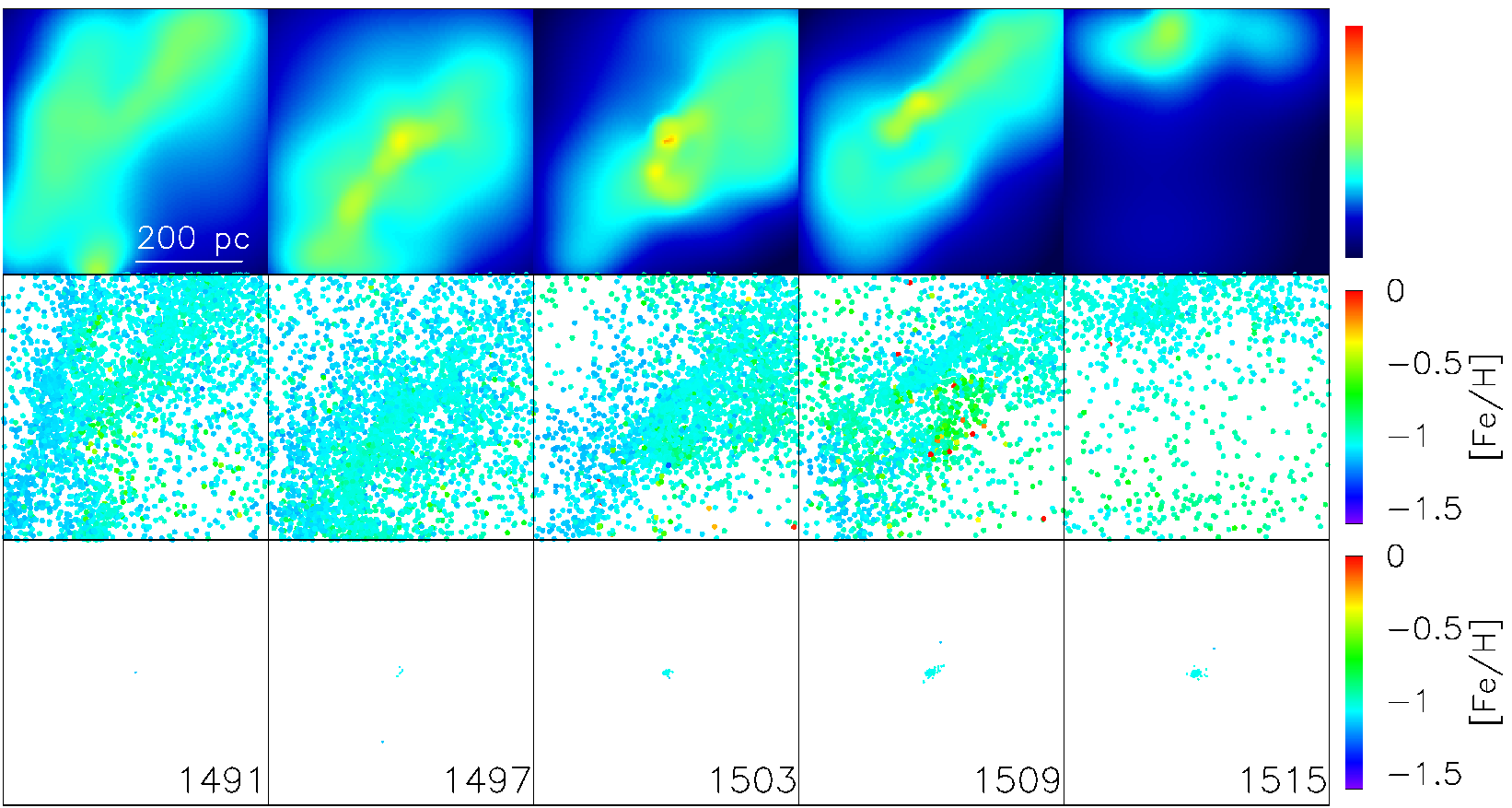}
 \end{center}
\caption{
Same figure as Figure~\ref{snapshots_ID3} but for ID 8 from $1491~{\rm Myr}$ to $1515~{\rm Myr}$.
}
\label{snapshots_ID8}
\end{figure}

Figure~\ref{snapshots_ID8} shows the snapshots of the formation process of the star cluster with ID 8, which does not have significant ${\rm [Fe/H]}$ abundance variations, from $1491~{\rm Myr}$ to $1515~{\rm Myr}$.
New stars begin to form from dense filament structures after $1497~{\rm Myr}$.
Although the 1P stars contaminate the surrounding gas by Type II SNe, the surrounding gas is expelled by Type II SNe before the contaminated gas falls into the cluster.
As a result, iron abundance in the cluster is predominantly homogeneous.

\begin{figure}
 \begin{center}
  \includegraphics[width=80mm]{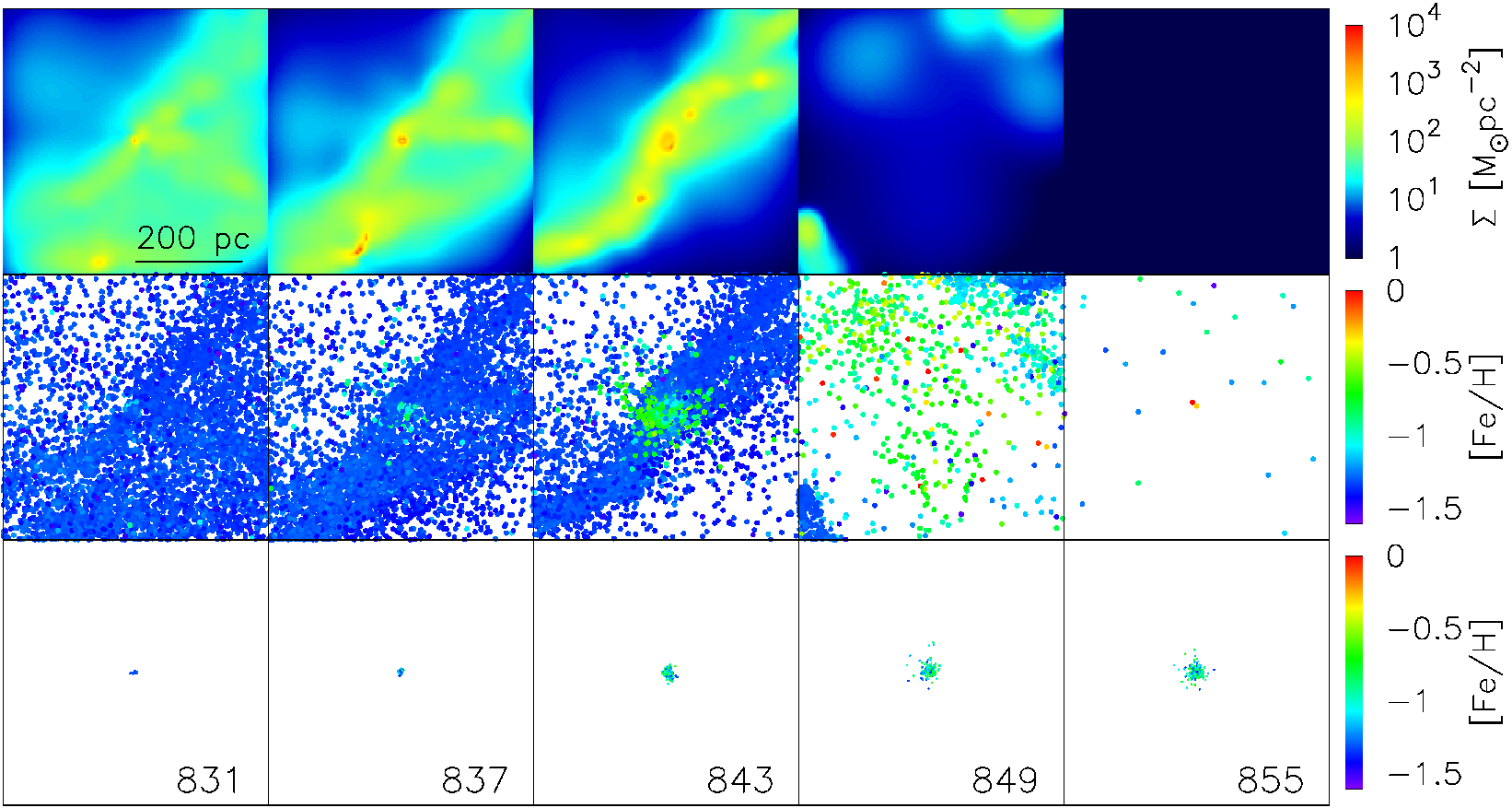}
 \end{center}
\caption{
Snapshots of the formation of the star cluster with ID 3 from $831~{\rm Myr}$ to $855~{\rm Myr}$.
The upper, middle, and bottom panels show the gas surface density, ${\rm [Fe/H]}$ of SPH particles, and ${\rm [Fe/H]}$ of stars constituting the star cluster, respectively.
The size of each panel is $500~{\rm pc}\times 500~{\rm pc}$.
}
\label{snapshots_ID3}
\end{figure}

\begin{figure}
 \begin{center}
  \includegraphics[width=80mm]{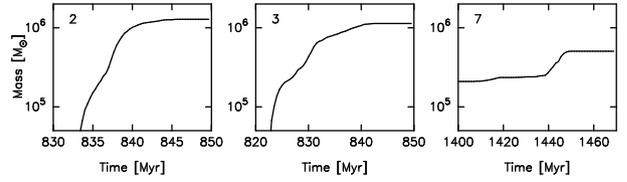}
 \end{center}
\caption{
Time evolution of masses of star clusters.
The left and middle panels show the mass evolution of the star clusters with ID 2 and ID 3, respectively.
These clusters are formed at $\sim 820~{\rm Myr}$ and $\sim 830~{\rm Myr}$, respectively.
The right panel shows the mass evolution of the star cluster with ID 7.
This cluster is formed before the galaxy merger.
}
\label{mass_ID3}
\end{figure}

On the other hand, there are some star clusters that have a non-negligible spread in [Fe/H].
Star clusters with IDs 2 and 3 have a metal-rich tail, reminiscent of the metallicity distribution function of the Galactic star cluster $\omega$ Centauri (e.g., \citet{2010ApJ...722.1373J,2021MNRAS.505.1645M}) which is a result of subsequent star formation events.
Figure~\ref{snapshots_ID3} and the middle panel of Figure~\ref{mass_ID3} show snapshots of formation of the star cluster with ID 3 and its time evolution of mass, respectively.
When the two gas disks firstly collide, gas is compressed at the collision interface, which produces a dense gas filament.
In the dense gas filament, a seed star cluster, of which mass is $\sim 5\times 10^5~{\rm M}_{\odot}$, begins to be formed at $\sim 823~{\rm Myr}$.
These stars are formed from gas with ${\rm [Fe/H]}\sim -1.3$ and represent the 1P stars.
These stars contaminate the surrounding gas by Type II SNe after $\sim 830~{\rm Myr}$, however at that time, Type II SN feedback is unable to expel the surrounding gas.
After $840~{\rm Myr}$, the contaminated gas falls into the cluster seed.
When the gas in the cluster becomes sufficiently dense to form stars, the 2P stars are formed from such gas.
The 2P stars are responsible for the higher [Fe/H] distribution components.
Finally, the star cluster's mass reaches $\sim 1\times 10^6~{\rm M}_{\odot}$.
At $\sim 849~{\rm Myr}$, strong energy injection to the surrounding gas by Type II SNe triggers radial outflow of the surrounding gas.
The depletion of gas around clusters quenches gas infall onto the cluster and halts any further star formation.
Similar evolution occurs for the star cluster with ID 2.

Whether or not a star cluster experiences self-enrichment depends on amount of the surrounding gas and the gravitational potential of the cluster.
As shown in Figure~\ref{snapshots_ID8} and Figure~\ref{snapshots_ID3}, the surrounding gas of the star cluster with ID 3 is more abundant than that of the star cluster with ID 8.
If the surrounding gas is abundant and falls into the star cluster before the gas outflow, the next generation of stars is formed in the cluster.
Therefore, star clusters with the deeper gravitational potential like the star cluster with ID 2 and 3 tend to obtain the surrounding gas since gas falls into such clusters faster.
Observationally, however, there is a trend that more massive clusters indeed are more likely to be Type II clusters (e.g., \citet{2019A&ARv..27....8G}), but this is not a rule (e.g., \citet{2021ApJ...923...22M}).

\begin{figure}
 \begin{center}
  \includegraphics[width=80mm]{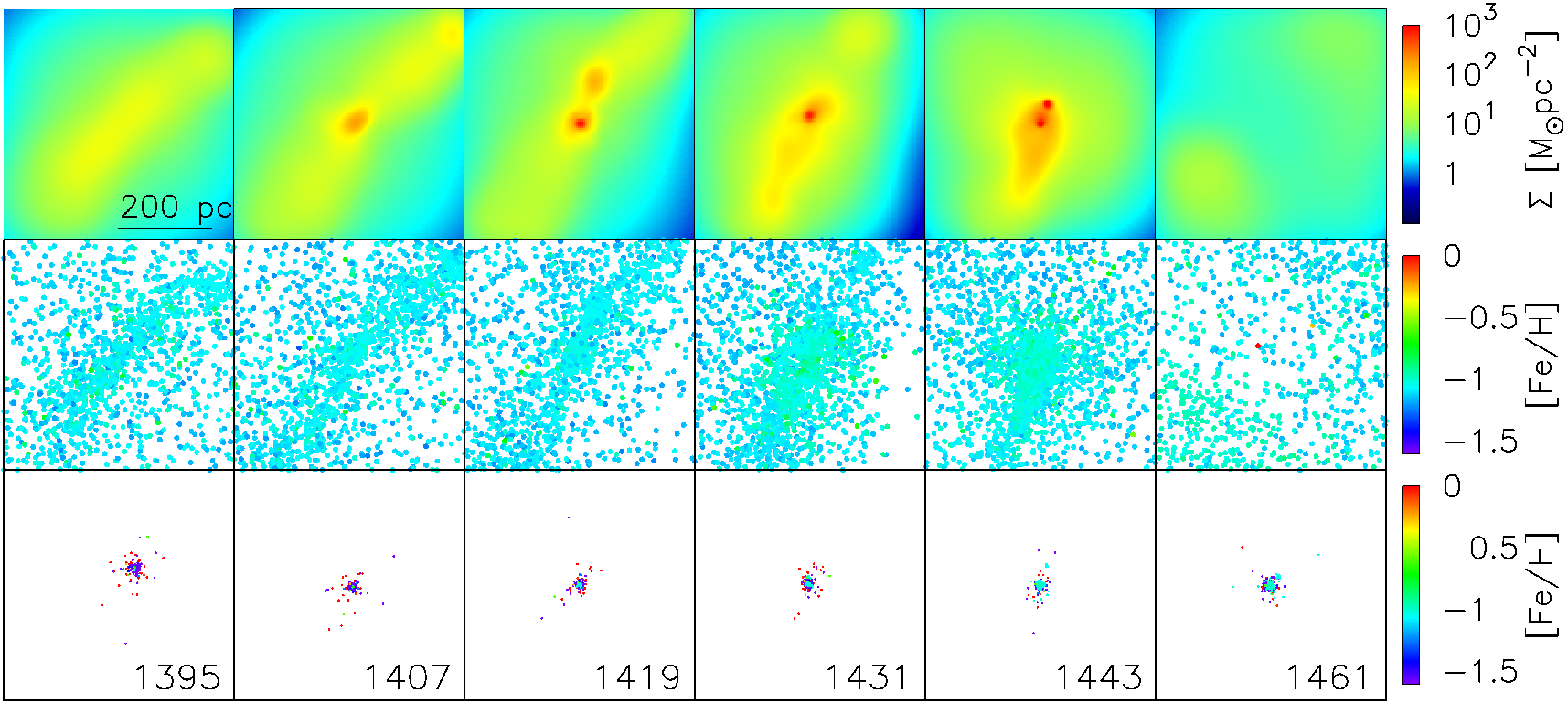}
 \end{center}
\caption{
Same as Figure~\ref{snapshots_ID3} but for the cluster with ID7 from $1395~{\rm Myr}$ to $1461~{\rm Myr}$.
}
\label{snapshots_ID7}
\end{figure}

The star cluster with ID 7 has a multimodal ${\rm [Fe/H]}$ distribution comprised of three components with peaks at ${\rm [Fe/H]}\sim -1.6$, $\sim -1.05$, and $\sim -0.95$.
Figure~\ref{snapshots_ID7} shows snapshots from $1395~{\rm Myr}$ to $1443~{\rm Myr}$ to illustrate this cluster's formation process.
A star cluster seed with a mass of $2.1 \times 10^5~{\rm M}_{\odot}$ is formed due to self-gravitational instability in the galactic disk before the galaxy-galaxy merger.
Most of the 1P stars have a ${\rm [Fe/H]}\sim -1.6$ as the stars are formed from pristine gas set at ${\rm [Fe/H]}\sim -1.6$.
The seed contains a small fraction of stars with ${\rm [Fe/H]}> 0.0$.
The seed cluster passes through a dense gas region at $\sim 1395~{\rm Myr}$ and captures a portion of this gas after $\sim 1407~{\rm Myr}$.
The 2P is created once the captured gas reaches a sufficient density for star formation, thereby increasing the star cluster mass by $2.6\times 10^4~{\rm M}_{\odot}$ from $1407~{\rm Myr}$ to $1418~{\rm Myr}$ (as shown in the right panel of Figure~\ref{mass_ID3}).
These stars have a metallicity abundance of $\rm{[Fe/H]}\sim -1.05$.
The 2P star particles explode as Type II SNe and contaminate the surrounding gas, raising the [Fe/H] abundance of the gas to $\sim -0.95$.
After the cluster captures this contaminated gas at $1430~{\rm Myr}$, the third population (3P) stars are formed within the cluster.

\begin{figure}
 \begin{center}
  \includegraphics[width=80mm]{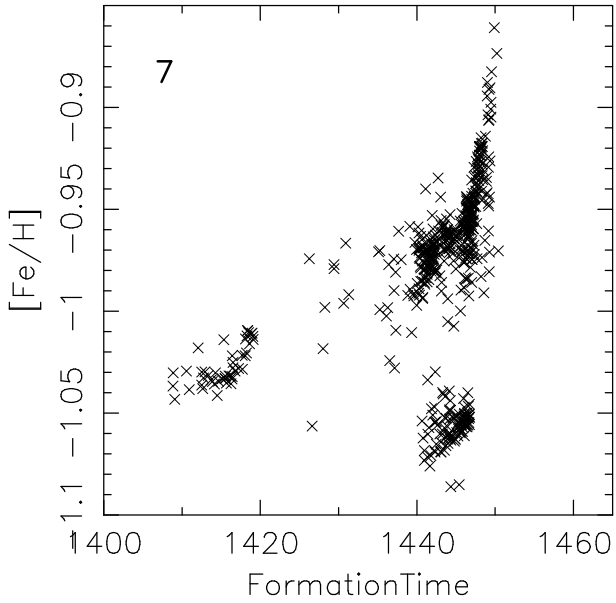}
 \end{center}
\caption{
Formation time v.s. [Fe/H] of the inner stars formed from $1400~{\rm Myr}$ to $1460~{\rm Myr}$ in the star cluster with ID 7.
}
\label{AgeFeH_ID7}
\end{figure}

In the gas surface density plot of the third panel from the left of Figure~\ref{snapshots_ID7} (at 1419 Myr), we can see a clumpy gas structure above the cluster.
This gas clump has evolved separately to ID 7 and is not contaminated by Type II SNe from the 2P star particles.
Therefore, its metallicity abundance is still ${\rm [Fe/H]}\sim -1.05$.
Since this clump falls into the cluster, the 3P stars are also formed from the gas with ${\rm [Fe/H]}\sim -1.05$.
Thus, these 3P stars have a large range in ${\rm [Fe/H]}$ as shown in Figure~\ref{AgeFeH_ID7}.
Because of the formation of the 3P stars, the stellar mass of the cluster finally increases to $5.1\times 10^5~{\rm M}_{\odot}$.
After $1455~{\rm Myr}$, energy injection into the surrounding medium by Type II SNe expels the surrounding gas out of the cluster.
 
In table~\ref{tab:clusters}, the standard deviation of the ${\rm [Fe/H]}$ abundance, $\sigma$, of the star cluster with ID 12 in ${\rm [Fe/H]}$ is $0.32$ although Figure~\ref{feh_distribution} shows that the star cluster has a single peak in ${\rm [Fe/H]}$ distributions.
This is because the cluster with ID 12 contains a minor fraction of stars with ${\rm [Fe/H]} \sim 0$.

\begin{figure}
 \begin{center}
  \includegraphics[width=80mm]{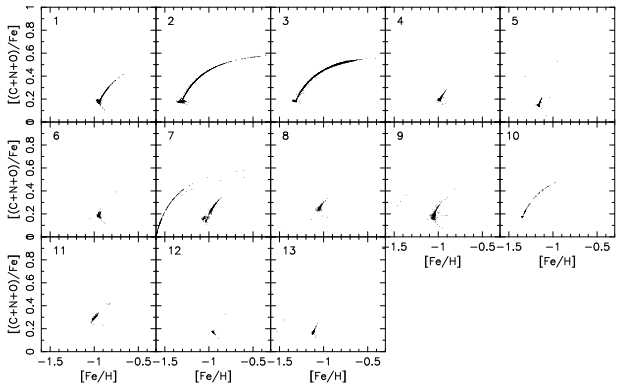}
 \end{center}
\caption{
${\rm [Fe/H]}$ v.s. ${\rm [(C+N+O)/Fe]}$ of the inner stars in each star cluster.
The upper left number in each panel shows the cluster ID.
}
\label{cno}
\end{figure}

\begin{figure}
 \begin{center}
  \includegraphics[width=80mm]{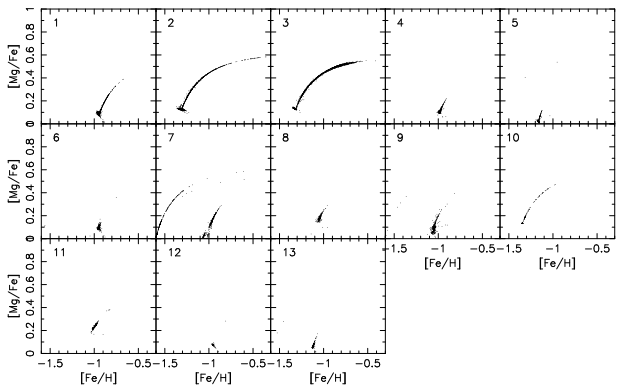}
 \end{center}
\caption{
Same as Figure~\ref{cno} but for ${\rm [Fe/H]}$ v.s. ${\rm [Mg/Fe]}$.
}
\label{mg}
\end{figure}

Both the kinematics and chemical abundances within the gas are strongly affected by Type II SNe.
Type II SNe contribute to the metal contamination of iron, but are the dominant source of $\alpha$ elements such as Na, Mg and Si (e.g., \citet{2020ApJ...900..179K}).
Figure~\ref{cno} and Figure~\ref{mg} show ${\rm [Fe/H]}$ v.s. ${\rm [(C+N+O)/Fe]}$ and ${\rm [Fe/H]}$ v.s. ${\rm [Mg/Fe]}$ of the inner stars in each cluster, respectively.
These figures show that there is a positive correlation between ${\rm [Fe/H]}$ and ${\rm [(C+N+O)/Fe]}$ and between ${\rm [Fe/H]}$ and ${\rm [Mg/Fe]}$.
${\rm [(C+N+O)/Fe]}$ and ${\rm [Mg/Fe]}$ approach asymptotically $\sim 0.6$ and $\sim 0.6$, which are the value predicted by the yield table from Type II SNe (see \citet{2017AJ....153...85S}), respectively.
These results indicate that clusters are mainly contaminated by Type II SNe.
Therefore, star clusters without variations of iron do not have variations of carbon, nitrogen, oxygen, and magnesium.
In the star clusters with metallicity variations, star formations occur continuously as shown in Figure~\ref{mass_ID3}.
The continued star formations from contaminated gas by Type II SNe result in a continuous increasing trend with metallicity rather than dichotomy in Figure~\ref{cno} and Figure~\ref{mg}.

 \subsection{Formation of a nuclear star cluster} \label{sec:nuclear_cluster}
 
The formation of star clusters during the second, third, and fourth encounters occurs in the central hundred pc region of the galaxies.
Therefore, 9 star clusters, except for three which formed during the first encounter, wander in the galactic central region after they form.
This process takes less than several hundred Myr due to dynamical friction and their falls lead to close encounters between the star clusters.

\begin{figure}
 \begin{center}
  \includegraphics[width=80mm]{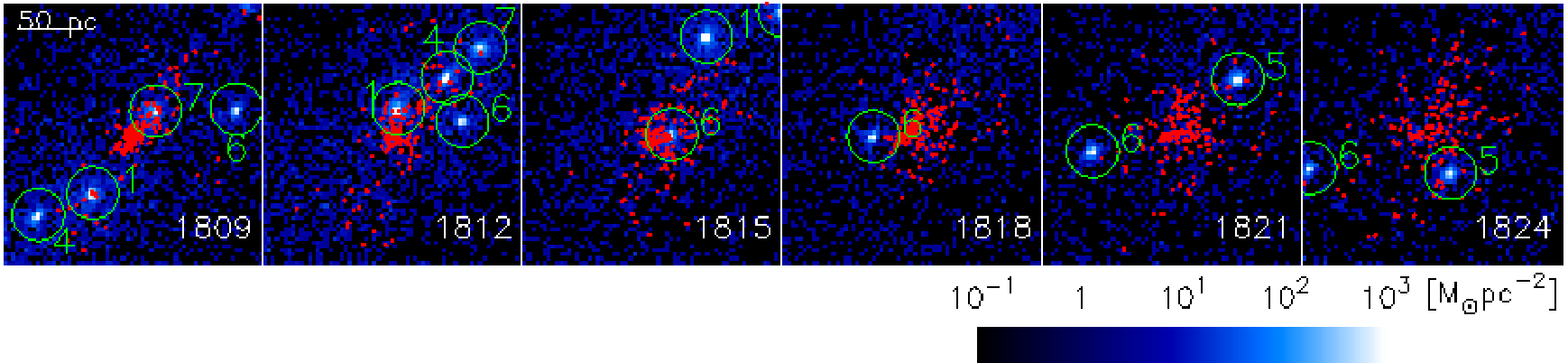}
 \end{center}
\caption{
Snapshots of close encounters between the cluster with ID 12 and the other clusters from $1809~{\rm Myr}$ to $1824~{\rm Myr}$.
The color map shows surface density of stars.
Red points are written on the color map and represent stars of the cluster with ID 12.
The green circles represent the star clusters with ID 1, 4, 5, 6, and 7.
The size of each panel is $200~{\rm pc} \times 200~{\rm pc}$.
}
\label{snapshots_ID12}
\end{figure}

Figure~\ref{snapshots_ID12} shows a snapshot from $1809~{\rm Myr}$ to $1824~{\rm Myr}$ of the dynamical evolution of the star cluster with ID 12, which has shallower gravitational potential due to close encounters between star clusters.
In this figure, red dots represent star particles that belonged to the star cluster with ID 12 at $1704~{\rm Myr}$.
After $1809~{\rm Myr}$, star particles from the cluster have already dispersed and created tidal tails due to frequent dynamical interactions with its surroundings.
These snapshots illustrate that the star cluster with IDs 1, 4, 5, 6, and 7 approaches the cluster with ID 12 within $\sim 20~{\rm Myr}$.
Due to its low binding energy, the multiple close encounters with the cluster with ID 12 further disperse its constituent star particles, resulting in its eventual tidal disruption after $1824~{\rm Myr}$.
The star cluster with ID 13 experiences a similar evolution to ID 12, and is also disrupted after $\sim 1920~{\rm Myr}$.
Searching for the tidal remnant of GCs is an important area of investigation within the Milky Way.
Fingerprints of GC formation from N enhanced 2P stars illustrate that GC dissolution is an important aspect of the build-up of the stellar halo of galaxies (e.g., \citet{2011A&A...534A.136M,2016ApJ...825..146M,2021MNRAS.500.5462H}).

Conversely, star clusters with ID 1, 4, 5, 6, 7, 8, and 9 have deeper gravitational potentials and are not tidally disrupted by close encounters between clusters.
Therefore, they sink into the galactic center due to the dynamical friction without tidal disruption and merge at the galactic center.
These mergers result in the formation of a nuclear star cluster.

\begin{figure}
 \begin{center}
  \includegraphics[width=80mm]{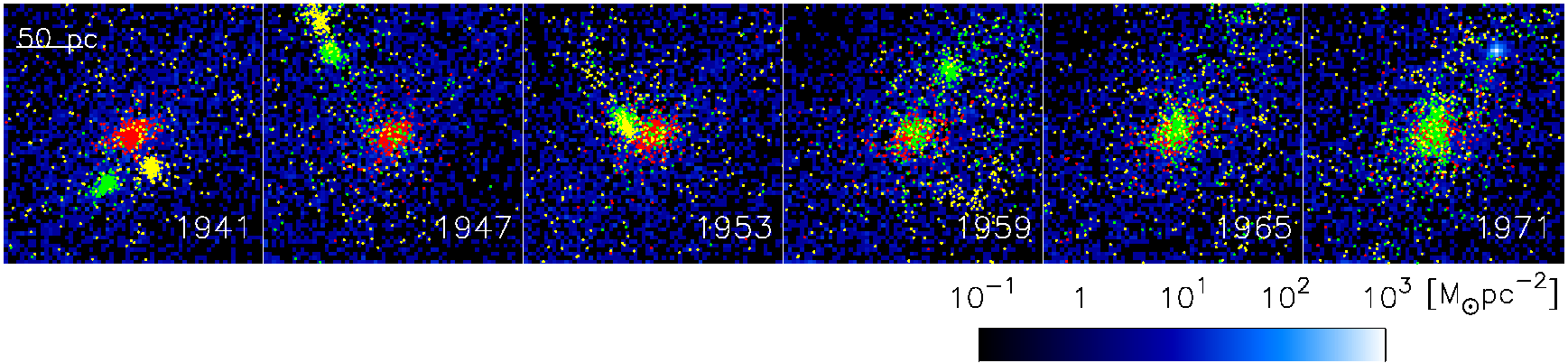}
 \end{center}
\caption{
Snapshots of mergers between the nuclear star cluster and star clusters with ID 4 and 9 from $1719~{\rm Myr}$ to $1731~{\rm Myr}$.
The color map shows surface density of stars.
Red, green, and yellow points are written on the map.
The red points represent star particles of the nuclear cluster.
The green and yellow points represent star particles of the clusters with ID 4 and 9, respectively.
The size of each panel is $200~{\rm pc} \times 200~{\rm pc}$.
}
\label{snapshots_ID4}
\end{figure}

Figure~\ref{snapshots_ID4} shows the snapshots of the merging process between the nuclear star cluster and star clusters with IDs 4 and 9 from $1941~{\rm Myr}$ to $1971~{\rm Myr}$.
Both clusters fall into the galactic center and approach the nuclear star cluster at $1940~{\rm Myr}$.
After the clusters experience some approaches and tidal interactions, they finally merge with the nuclear star cluster at $1953~{\rm Myr}$ and $1971~{\rm Myr}$, respectively.
As a result, the nuclear star cluster accumulates a part of the star clusters with IDs 4 and 9, while tidal interactions also scatter stars of the nuclear star cluster.

The nuclear star cluster is composed of stars originating from several star clusters. It has a final mass of $1.75\times 10^6~{\rm M}_{\odot}$ at a radius of $20~{\rm pc}$ after $2499~{\rm Myr}$.
The mass of stars that originated from the cluster with ID 1 accounts for $87$~\% of the total mass of the nuclear star cluster as this cluster was the most massive and had the deepest gravitational potential.
The nuclear star cluster contains the other components that originated from star clusters with IDs 4, 5, 6, 7, and 9.
Their mass fractions relative to the total mass of the nuclear star cluster are $6.4$~\%, $1.9$~\%, $0.5$~\%, $3.3$~\%, and $0.4$~\%, respectively.
If instead, we adopt $100~{\rm pc}$ for the truncated radius of the nuclear star cluster, its mass becomes $2.0\times 10^6~{\rm M}_{\odot}$ and the mass contributions of the stars originated from star clusters with ID 1, 4, 5, 6, 7, 8, 9, 11, 12, and 13 to the nuclear star cluster are $51$~\%, $11$~\%, $7.7$~\%, $4.1$~\%, $9.4$~\%, $0.2$~\%, $3.5$~\%, $0.09$~\%, $0.4$~\%, and $0.1$~\%, respectively.
This result indicates that the other star clusters, except the star cluster with ID 1, are scattered by tidal forces when they sink into the galactic center.
We should note that there is no primordial nuclear star cluster prior to a galaxy merger.
The nuclear star cluster can be assembled by mergers of star clusters in the case without a primordial nuclear star cluster.

\begin{figure}
 \begin{center}
  \includegraphics[width=80mm]{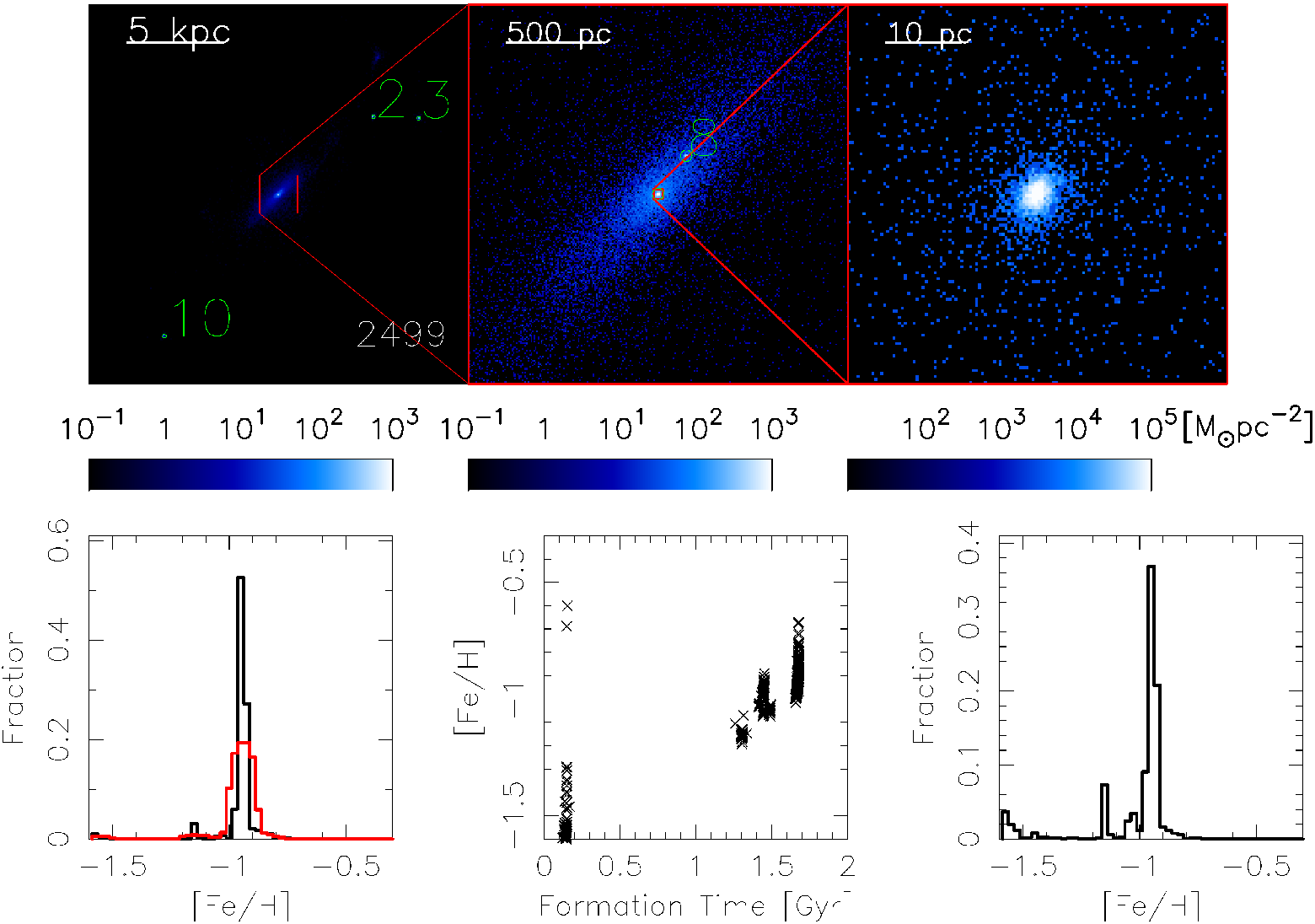}
 \end{center}
\caption{
The upper panels show the surface density map of newly formed stars.
The middle and right panels zoom in on the red squares of the right panel and the middle panel, respectively.
In the panels, the IDs are attached to the clusters.
The right panel shows the nuclear star cluster.
The panel sizes of the left, middle, and right are $20~{\rm kpc}\times 20~{\rm kpc}$, $2~{\rm kpc}\times 2~{\rm kpc}$, and $50~{\rm pc}\times 50~{\rm pc}$.
The lower left and middle panels show distributions of stellar populations in [Fe/H] and [Fe/H] versus star formation time of the nuclear star cluster.
The lower right panel shows distributions of stellar populations in [Fe/H] in the galactic central $100~{\rm pc}$.
}
\label{nuclear_sc}
\end{figure}

The lower left panel of Figure~\ref{nuclear_sc} shows distributions of stellar populations of the nuclear star cluster in ${\rm [Fe/H]}$.
The main distribution ranges from ${\rm [Fe/H]}=-1.0$ to ${\rm [Fe/H]}=-0.9$.
To make a more realistic comparison with observations, the red line in the panel shows the distributions of stellar populations by considering observational uncertainties.
Here, we give each particle a Gaussian metallicity spread with $\sigma \sim 0.05$.
The observational uncertainties widen the distribution, likening it more to observations of Galactic star clusters (see \S \ref{sec:discussion3}).
The lower central panel of Figure~\ref{nuclear_sc} shows the ${\rm [Fe/H]}$ abundance compared to the star formation time of the nuclear star cluster.
This panel shows that the nuclear star cluster consists of four main generations of stars and the stellar ages reflect the time of the second, third, and fourth encounters.
The lower right panel of Figure~\ref{nuclear_sc} shows the ${\rm [Fe/H]}$ distribution of the stellar populations of the galactic central $100~{\rm pc}$ region.
By allowing for a larger cut-off radius, the different sub-populations become more evident in the distribution.
The dominant population at ${\rm [Fe/H]}$ is maintained, however, additional peaks at ${\rm [Fe/H]}\sim -1.6$ and $\sim -1.15$ appear.
We discuss the observational implications of this in \S \ref{sec:discussion3}.

 \subsection{Distributions of star clusters in a merger remnant} \label{sec:remnant}

The upper panels of Figure~\ref{nuclear_sc} show the stellar distributions in the merger remnant after $2499~{\rm Myr}$.
Each panel gives a progressively zoomed-in view of the nuclear star cluster.
Green circles denote the locations of the five star clusters with IDs 1 (the main component of the nuclear star cluster), 2, 3, 8 and 10.

In the galactic central region, only two star clusters with IDs 1 and 8 are visible as many of the identified clusters have already merged to create the nuclear star cluster as described in \S \ref{sec:nuclear_cluster}.
In the outer galactic region, three clusters with IDs 2, 3, and 10 are observable.
All of them form during the first encounter and rotate around the center of the merger remnant.
The apocenter distances of the clusters are $10~{\rm kpc}$, $12~{\rm kpc}$, and $11~{\rm kpc}$, respectively.
The reason for their large apocenter distances is due to the dense gas filament that formed at the collision interface dissipating the kinetic energy of the gas and decreasing the velocity.
Consequently, the velocities of star clusters formed in this filament are also slow.
On the other hand, the galaxies continue to move without such deceleration after the first encounter.
After the encounter, the maximum distance between the star clusters and the galactic centers reaches $\sim 10~{\rm kpc}$.
Since the star clusters fall into the galaxy without a large tangential velocity, their orbits become highly eccentric.
The pericenter distances of these clusters are $0.21~{\rm kpc}$, $0.56~{\rm kpc}$, and $1.77~{\rm kpc}$, respectively.

\section{Discussions}  \label{sec:discussion}

\subsection{Formation mechanism of Type I and II globular clusters}

Our simulation has revealed that massive star clusters with and without large ($>0.1~{\rm dex}$) [Fe/H] spreads are formed during a dwarf-dwarf merger process.
Abiding by the conventions set by \citet{Milone+2017}, Type II GCs contain populations of stars with anomalous chemical abundance variations in elements heavier than iron.
Conversely, Type I clusters contain multiple populations in terms of their C, N, Na, O, Mg and Al abundances.
However, resolution of our simulation is insufficient to demonstrate a Na-O or a Mg-Al anti-correlation.
Therefore, we take a more conservative approach and state that these clusters are candidate Type I GCs.

Several works have investigated the process of GC formations via simulations \citep{2019MNRAS.489.3269C,2021MNRAS.507..834M,2021MNRAS.500.4578M}.
In our simulation, Type II star clusters with ${\rm [Fe/H]}$ variations are created by the following process.
Firstly, a star cluster is formed from dense gas produced in the merging process of dwarf galaxies.
These 1P stars explode as Type II SNe, contaminating the surrounding gas, and increasing the overall metallicity.
The energy injection from Type II SNe is unable to completely expel the surrounding gas and the contaminated gas falls into the cluster and forms the 2P stars with a higher ${\rm [Fe/H]}$ abundance.
As a result, a star cluster with multiple stellar populations and ${\rm [Fe/H]}$ variations is formed.

On the other hand, star clusters without significant ${\rm [Fe/H]}$ abundance variations do not capture gas contaminated by SNe.
Type II SNe expel the surrounding gas before the contaminated gas falls into the cluster.
Consequently, the 2P stars does not form in the cluster, resulting in a predominantly homogeneous iron abundance.

The star cluster with ID 7 consists of three population stars.
The first population is formed before the dwarf galaxy merger captures dense gas, and the second and third are formed from captured gas.
Based on photometry, \citet{Piotto+2007} observed a triple main sequence in the Galactic GC NGC 2808, implying that it harbours three populations.
When comparing the distribution of ${\rm [Fe/H]}$ from \citet{Carretta2015} for NGC 2808, the multi-modal features seen in the ${\rm [Fe/H]}$ distribution from the cluster with ID 7 are not present.
The cluster M 22 is famous for its two iron populations \citep{2011A&A...532A...8M}, but the difference is not this extreme, and the iron-rich population in M 22 cannot be further divided into multiple subpopulations as it can be for the cluster with ID7.

\subsection{Comparison with previous simulations of galaxy mergers} \label{sec:discussion2}

We compare our simulation of a dwarf galaxy merger with previous simulations of more massive merging galaxies.
\citet{2012ApJ...746...26M} performed galaxy merger simulations with a mass of $6.3\times 10^9~{\rm M}_{\odot}$ for each galactic disk which is more massive than our dwarf galaxy model.
The simulations illustrated that strong gas inflow takes place in the merger process, which is expected to power active galactic nuclei (AGN) activity and induce the growth of a galactic central supermassive black hole (SMBH).
Furthermore, such gas inflow triggers a nuclear starburst.
The nuclear starburst heats the surrounding dusts, thereby increasing infrared luminosity emitted from a merger remnant.
The luminosity reaches $10^{11}~{\rm L}_{\odot}$ which is comparable to that of a luminous infrared galaxy (LIRG).

On the contrary, our simulation has shown that gas is expelled from our dwarf galaxies due to the strong energy injection by Type II SNe after the fourth encounter.
The radial outflow is due to the shallower galactic potential of the dwarf galaxies and as a result, gas fueling to the galactic central region does not occur in the dwarf-dwarf merging process, unlike massive galaxy mergers.
The gas outflow also terminates star formation in the remnant.
Recent observations have shown that some dwarf-dwarf merging galaxies possess double AGN \citep{2024MNRAS.tmp..271M}.
Thus such AGN activities may be induced before the gas outflow.

The previous merger simulations of massive galaxies have revealed that abundant gas exists in the galactic central region due to strong inflows.
A nuclear starburst contaminates the gas in the central region of a merger remnant.
When star clusters pass through the contaminated dense gas region, they capture such gas due to the gravitational focusing effect.
As a result, an additional population of stars is formed in the star cluster \citep{2019PASJ...71...19M}.
These clusters have large age spreads since the dense gas is captured several hundred Myr after the clusters are formed.
On the other hand, in the dwarf-dwarf merger process, star clusters with large age spreads are not formed except for ID 7 which is formed before the first encounter of galaxies.
This is attributed to gas depletion in the merger remnant due to strong outflow by Type II SNe.
\citet{2019PASJ...71...19M} do not consider metal contamination in their simulations and cannot comment on the self-metal enrichment of star clusters such as the star clusters with ID 2, 3, or 7.
We will investigate the self-metal enrichment of star clusters in massive galaxy mergers in future works.

In our dwarf galaxies merger simulation, star clusters fall into the galactic center due to dynamical friction.
These clusters merge in the galactic central region, thereby creating a nuclear star cluster.
Previous simulations of massive galaxy mergers have revealed that some massive star clusters are formed in the galactic central region within a few hundred Myr and merge with the galactic core \citep{2012ApJ...746...26M}.
These results suggest that nuclear star clusters or cores in merger remnants generally experience mergers of star clusters that fall into the galactic center.
Such nuclear star clusters naturally include multiple populations differing in ${\rm [Fe/H]}$ abundances.

\subsection{Multiple populations in a nuclear star cluster} \label{sec:discussion3}

Previous studies have suggested two formation scenarios of a nuclear star cluster.
One is mergers of GCs (e.g., \citet{1975ApJ...196..407T,2011ApJ...729...35A,2014ApJ...785...71G,2022A&A...667A.101F}), and another is in situ star formations in the galactic nuclear region (e.g., \citet{2006ApJ...642L.133B,2015ApJ...812...72A}).
Our simulation is the first simulation which demonstrates that a nuclear star cluster is created by mergers of star clusters formed in the merging process of dwarf galaxies, which supports the former scenario.
Additionally, the nuclear star cluster would not contain a young stellar population in the merger remnant since star formations hardly occur in the nuclear region due to gas outflow by Type II SNe.
On the other hand, previous merger simulations of more massive galaxies have shown that gas inflow into the galactic center occurs during the merging of two massive galaxies as described in \S \ref{sec:discussion2}.
The gas inflow is expected to trigger star formations in the nuclear star cluster.
Thus, in situ star formations would contribute to formation of a nuclear star cluster in massive galaxies, which supports the latter scenario.
These results agree with observations which indicates that nuclear star clusters are formed predominantly by mergers of GCs in less massive galaxies whereas in situ star formations in the galactic nuclear region contribute to formation of a nuclear star cluster in massive galaxies \citep{2022A&A...667A.101F}.

Recent observations have revealed that GC candidates are concentrated in the central region of a dwarf-dwarf merging galaxy \citep{2023A&A...671L...7R}.
Such concentration is expected to be triggered by the dynamical friction.
Our simulation suggests that these candidates continue to fall into the center of the merger remnant, leading to the formation of a nuclear star cluster.
Star clusters with deeper gravitational potentials can fall into the galactic center without tidal disruption, while the clusters with shallower potential tend to be tidally disrupted.

We compare the nuclear star cluster formed in our simulation to the massive Galactic star cluster $\omega$ Cen which, although usually given the classification of a GC, has been hypothesized to originally be a galactic core of a dwarf galaxy \citep{Freeman+1993,2003MNRAS.346L..11B}.
$\omega$ Cen is one of the most well-studied clusters in the Milky Way and is unique in terms of the population of Galactic GCs (e.g., \citet{2001A&A...369...87G,2004ARA&A..42..385G,2011A&A...534A..72G}).
Observations have indicated that the $\omega$ Cen is extremely massive ($\sim 4\times 10^6~{\rm M}_{\odot}$ (e.g., \citet{2015ApJ...812..149W})) and has been suggested to contain as many as 15 sub-populations \citep{2017ApJ...842....7B}.
$\omega$ Cen consists of stars with iron variations with a metallicity range of $-2.2 < {\rm [Fe/H]} < -0.6$ \citep{2010ApJ...722.1373J}.

In our simulation, the nuclear star cluster is created through the assembly of nine star clusters formed by various close encounters.
The nuclear star cluster has a mass of $\sim 10^6~{\rm M}_{\odot}$ and consists of four population stars which have a metallicity range of $-1.6 < {\rm [Fe/H]} < -0.8$.
Its mass, populations, and metallicity range seem to be lower than the present day values of $\omega$ Cen.

After the fourth encounter, the gas is completely expelled from the two galaxies and star clusters can no longer form.
The availability of star-forming gas is one parameter that drives the diversity of our candidate globular clusters.
Thus, if these clusters were to fall into the galactic center and create a nuclear star cluster, this would increase the number of distinct subpopulations within the nuclear star cluster.
In addition, if the gas fraction is higher in the dwarf galaxies before a merger, we predict that a larger number of star clusters would be formed, and the ISM of the galaxies would become more metal-rich.
Therefore, we find that orbital and collision parameters are important parameters for generating the mass and chemical abundance profile of the nuclear star cluster.

\section{Conclusion}  \label{sec:conclusion}

In this work, we have performed a high-resolution numerical simulation of a dwarf galaxy merger.
We summarize our main conclusions as follows:
\begin{itemize}
\item Our simulation of a dwarf-dwarf galaxy merger forms 13 star clusters that can be separated into two types.
We liken these to observational works which divide clusters into those that do not have statistically significant iron abundance variations (Type I), and clusters which do show evidence Fe abundance variations (Type II).
\item The star clusters with ${\rm [Fe/H]}$ abundance variations are created via the following process:
A star cluster, which consists of 1P stars, is formed from dense gas during a galaxy merger.
Type II SNe ejecta from the 1P contaminates the surrounding gas, however, the energy injection is unable to expel the surrounding gas completely.
The contaminated gas falls into the cluster and the 2P stars are formed from such gas in the cluster.
\item Nine star clusters fall into the galactic center due to dynamical friction.
Although a portion of the clusters are disrupted by tidal interactions between clusters, the others merge at the galactic central region, thereby forming a nuclear star cluster.
As a result, the nuclear star cluster has widespread ${[\rm Fe/H]}$ variations.
\item In the dwarf-dwarf merger remnant, there are three star clusters in the galactic outer region.
These clusters are formed in the first encounter and rotate around the merger remnant.
Their orbits are highly eccentric.
On the other hand, there are few star clusters in the galactic central region since most have fallen into the galactic center due to the dynamical friction within several hundred Myr.
\end{itemize}

In this paper, we perform a simulation with one dwarf galaxy and collision model.
The galaxy mergers take place randomly in a hierarchical galaxy formation process.
Therefore, simulations with various galaxy models and collision parameters are needed to understand the properties of massive star clusters in a dwarf galaxy merger process further.
In future works, we will perform simulations with various mass ratios, gas fractions, orbits (the pericenter and apocenter distance), and collision parameters of merging galaxies.

\bigskip
We are grateful to the anonymous referee for giving us constructive and useful comments. 
This work was supported by JSPS KAKENHI Grant Number 21K03621.
Numerical computations were carried out on Cray XC50 at Center for Computational Astrophysics (CfCA), National Astronomical Observatory of Japan, and numerical analyses were also carried out on the analysis servers at CfCA.


\end{document}